\documentclass[twocolumn,showpacs,preprintnumbers,amsmath,amssymb]{revtex4}

\usepackage{graphicx}
\usepackage{dcolumn}
\usepackage{bm}
\usepackage{psfig}
\usepackage{subfigure}

\newcommand{\goodvgap}{\vspace{-0.5cm}}

\begin{document}

\title{Quantum dot single photon sources: Prospects for applications in linear optics quantum
information processing}

\author{A.~Kiraz$^{1}$, M.~Atat\"ure$^{2,3}$, and A. Imamo\=glu$^{2,3,4}$}

\address{$^1$ Department Chemie, Ludwig-Maximilians Universit{\"a}t
M{\"u}nchen, Butenandtstr. 11, D-81377 Munich, Germany\\
$^2$ Quantenelektronik, ETH-H\"onggerberg, HPT G12, CH-8093
Zurich, Switzerland \\
$^3$ IV. Physikalisches Institut, Universitat Stuttgart,
Pfaffenwaldring 57, D-70550 Stuttgart, Germany\\
$^4$ Faculty of Engineering and Natural Sciences, Sabanci
University, Istanbul, Turkey}
\date{\today}


\begin{abstract}
An optical source that produces single photon pulses on demand has
potential applications in linear optics quantum computation,
provided that stringent requirements on indistinguishability and
collection efficiency of the generated photons are met. We show
that these are conflicting requirements for anharmonic emitters
that are incoherently pumped via reservoirs. As a model for a
coherently pumped single photon source, we consider
cavity-assisted spin-flip Raman transitions in a single charged
quantum dot embedded in a microcavity. We demonstrate that using
such a source, arbitrarily high collection efficiency and
indistinguishability of the generated photons can be obtained
simultaneously with increased cavity coupling. We analyze the role
of errors that arise from distinguishability of the single photon
pulses in linear optics quantum gates by relating the gate
fidelity to the strength of the two-photon interference dip in
photon cross-correlation measurements. We find that performing
controlled phase operations with error $< 1 \%$ requires
nano-cavities with Purcell factors $F_P \ge 40$ in the absence of
dephasing, without necessitating the strong coupling limit.
\end{abstract}


\maketitle

\section{Introduction}
A significant fraction of key experiments in the emerging field of
quantum information science \cite{NC}, such as Bell's inequality
violations \cite{aspect2}, quantum key distribution
\cite{QKD,QKD2} and quantum teleportation \cite{QT} have been
carried out using single photon pulses and linear optical elements
such as polarizers and beam splitters. However, it was generally
assumed that in the absence of photon-photon interactions, the
role of optics could not be extended beyond these rather limited
applications. Recently, Knill, Laflamme, and Milburn have shown
theoretically that efficient linear optics quantum computation
(LOQC) can be implemented using on-demand indistinguishable
single-photon pulses and high-efficiency photon-counters
\cite{KLM01}. This unexpected result has initiated a number of
experimental efforts aimed at realizing suitable single-photon
sources. Impressive results demonstrating a relatively high degree
of indistinguishability and collection efficiency have been
obtained using a single quantum dot embedded in a microcavity
\cite{santori2002}. Two-photon interference has also been observed
using a single cold atom trapped in a high-Q Fabry-Perot
cavity~\cite{rempe2003}. A necessary but not sufficient condition
for obtaining indistinguishable single photons on demand is that
the cavity-emitter coherent coupling strength ($g$) exceeds the
square root of the product of the cavity ($\kappa_{cav}$) and
emitter ($\gamma$) coherence decay rates. When the emitter is
spontaneous emission broadened and the cavity decay dominates over
other rates, this requirement corresponds to the Purcell regime
($g^2/\kappa_{cav}\gamma>1$).

\begin{figure}
\begin{center}
  \centerline{\psfig{figure=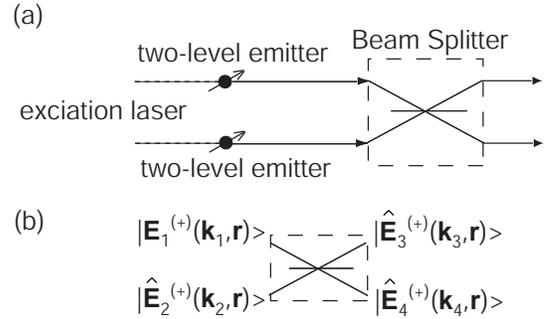,height=1.7in}}
  \caption{(a) Configuration assumed in the analysis of two-photon interference: Two
  independent identical single photon sources excited by the same laser field. (b) Input and output fields of the beam splitter.}
\goodvgap
\label{simplified_scheme}
\end{center}
\end{figure}
In this paper, we identify the necessary and sufficient conditions
for generation of single photon pulses with an arbitrarily high
collection efficiency and indistinguishability. While our results
apply to all single-photon sources based on two-level emitters,
our focus will be on quantum dot based devices. First, we show
that single photon sources that rely on incoherent excitation of a
single quantum dot (through a reservoir) cannot provide high
collection efficiency and indistinguishability, simultaneously. To
achieve this goal, the only reservoir that the emitter couples to
has to be the radiation field reservoir that induces the cavity
decay. We show that a source based on cavity-assisted spin-flip
Raman transition satisfies this requirement and can be used to
generate the requisite single-photon pulses in the Purcell regime.
This analysis is done in section~\ref{sec_indist} where we
calculate the degree of interference (indistinguishability) of two
photons and the theoretical maximum collection efficiency, as a
function of the cavity coupling strength, laser pulsewidth, and
emitter dephasing rate for different single photon sources.

Interference of two single-photon pulses on a beam-splitter plays
a central role in all protocols for implementing indeterministic
two-qubit gates, which are in turn key elements of linear optics
quantum computation schemes \cite{KLM01}. Observability of
two-photon interference effects naturally requires that the two
single-photons arriving at the two input ports of the
beam-splitter be indistinguishable in terms of their pulsewidth,
bandwidth, polarization, carrier frequency, and arrival time at
the beam-splitter. The first two conditions are met for an
ensemble of single-photon pulses that are Fourier-transform
limited: this is the case if the source (single atom or quantum
dot) transition is broadened solely by spontaneous emission
process that generates the photons. While the radiative lifetime
(i.e. the single-photon pulsewidth) of the emitter does not affect
the observability of interference, any other mechanism that allows
one to distinguish the two photons will. A simple example that is
relevant for quantum dot single photon sources is the uncertainty
in photon arrival (i.e. emission) time arising from the random
excitation of the excited state of the emitter transition: if for
example this excited state is populated by spontaneous phonon
emission occuring with a waiting time of $\tau_{relax}$, then the
{\sl starting time} of the photon generation process will have a
corresponding time uncertainty of $\sim \tau_{relax}$. We refer to
this uncertainty as time-jitter. Since the information about the
photon arrival time is now carried by the phonon reservoir, the
interference will be degraded.

Even though the role of single-photon loss on linear optics
quantum computation has been analyzed~\cite{KLM01}, there has been
to date no analysis of gate errors arising from distinguishability
of single photons. To this end, we first note that while various
sources of distinguishability can be eliminated, the inherent
jitter in photon emission time remains as an unavoidable source of
distinguishability. Hence, in section~\ref{LOQC_sec}, we analyze
the performance of a linear-optics-controlled phase gate in the
presence of time-jitter and relate the gate fidelity to the degree
of indistinguishability of the generated photons, as measured by a
Hong-Ou-Mandel~\cite{hong1987} type two-photon interference
experiment.

\section{Maximum collection efficiency and indistinguishability of
photons generated by single photon sources}\label{sec_indist} In
this section we first develop the general formalism for
calculating a normalized measure of two-photon interference based
on the projection operators of a two-level emitter. We then
compare and contrast the case where the emitter is pumped via
spontaneous emission of a photon or a phonon from an excited
state, i.e. an incoherently pumped single photon source, to the
case where single photon pulses are generated by cavity-assisted
spin-flip Raman scattering, i.e. coherently pumped single photon
source.

Previous analysis of two-photon interference among photons emitted
from single emitters were carried out for two-level systems driven
by a cw laser field~\cite{H.Fearn:89,bylander2003}. In contrast,
we treat the pulsed excitation, and analyze currently available
single photon sources based on two and three-level emitters. We
note that extensive analysis of two-photon interference phenomenon
was carried out for twin photons generated by parametric down
conversion~\cite{ghosh1986,hong1987,shih1988,steinberg1992}, and
single photon wave-packets~\cite{legero2003}, without considering
the microscopic properties of the emitter.

\subsection{Calculation of the degree of two-photon interference}\label{sec_normalization}
We consider the experimental configuration depicted in
Figure~\ref{simplified_scheme}(a). Two general independent
identical two-level emitters are assumed to be excited by the same
laser. We assert no further assumptions on two-level emitters;
they are considered to be light sources that exhibit perfect
photon antibunching. Single photons emitted from the two-level
emitters are coupled to different inputs of a beam splitter which
is equidistant from both sources. In the ideal scenario where the
input channels are mode-matched and the incoming photons have
identical spectral and spatial distributions, two-photon
interference reveals itself in lack of coincidence counts among
the two output channels. This bunching behavior is a signature of
the bosonic nature of photons.

Recent demonstration of two-photon interference using a single
quantum dot single photon source relied on a similar scheme based
on a Michelson interferometer~\cite{santori2002}. In this
experiment, the interferometer had a large path length difference
between its two branches. Such a difference, in excess of single
photon coherence length, provided the interference among photons
subsequently emitted from the same source. Two-photon interference
in this experiment is quantitatively similar to interference
obtained among photons emitted by two different identical sources.

Input-output relationships for single mode photon annihilation
operators in the beam splitter (Fig.~\ref{simplified_scheme}(b))
are defined by the unitary operation
\begin{eqnarray}
    \left[
    \begin{array}{c}
        \hat{a}_3(\omega)\\
        \hat{a}_4(\omega)
    \end{array}
    \right]
=
    \left[
    \begin{array}{cc}
    \cos\xi & -e^{-i\phi}\sin\xi\\
    e^{i\phi}\sin\xi & \cos\xi
    \end{array}
    \right]
    \left[
    \begin{array}{c}
        \hat{a}_1(\omega)\\
        \hat{a}_2(\omega)
    \end{array}
    \right]\,.
    \label{k_beam_split}
\end{eqnarray}
$\hat{a}_1(\omega)$, $\hat{a}_2(\omega)$, $\hat{a}_3(\omega)$, and
$\hat{a}_4(\omega)$ represent single mode photon annihilation
operators in channels $\mathbf{k_1}$, $\mathbf{k_2}$,
$\mathbf{k_3}$, and $\mathbf{k_4}$ respectively. $\mathbf{k_1}$,
$\mathbf{k_2}$, $\mathbf{k_3}$, and $\mathbf{k_4}$ have identical
amplitudes and polarizations while satisfying the momentum
conservation. We will abbreviate the unitary operation in the beam
splitter as $u(\textbf{B}_{\xi,\phi})$.

Assuming that $u(\textbf{B}_{\xi,\phi})$ is constant over the
frequency range of consideration, Eq.~(\ref{k_beam_split}) can be
Fourier transformed to reveal
\begin{eqnarray}
    \left[
    \begin{array}{c}
        \hat{a}_3(t)\\
        \hat{a}_4(t)
    \end{array}
    \right]
= u(\textbf{B}_{\xi,\phi})
    \left[
    \begin{array}{c}
        \hat{a}_1(t)\\
        \hat{a}_2(t)
    \end{array}
    \right]\,.
    \label{t_beam_split}
\end{eqnarray}
$\hat{a}_1(t)$, $\hat{a}_2(t)$, $\hat{a}_3(t)$, and $\hat{a}_4(t)$
now represent time dependent photon annihilation operators.

Coincidence events at the output of the beam splitter are
quantified by the cross-correlation function between channels 3
and 4 which is given by
\begin{eqnarray}
 G^{(2)}_{34}(t,\tau)&=&\langle
\hat{a}_3^\dag(t)\hat{a}_4^\dag(t+\tau)\hat{a}_4(t+\tau)\hat{a}_3(t)\rangle \,,
\label{g2_unnorm}
\end{eqnarray}
\begin{eqnarray}
g^{(2)}_{34}(t,\tau) &=& \frac{G^{(2)}_{34}(t,\tau)}{\langle
\hat{a}_3^\dag(t)\hat{a}_3(t)\rangle\langle
\hat{a}_4^\dag(t+\tau)\hat{a}_4(t+\tau)\rangle} \,, \label{g2_nor}
\end{eqnarray}
in its unnormalized ($G^{(2)}_{34}(t,\tau)$) and normalized
($g^{(2)}_{34}(t,\tau)$) form. By substitution of
Eq.~(\ref{t_beam_split}) in (\ref{g2_unnorm}),
$G^{(2)}_{34}(t,\tau)$ is expressed as
\begin{eqnarray}
G^{(2)}_{34}(t,\tau) & = & \sin^4\xi \langle
\hat{a}_2^\dag(t)\hat{a}_1^\dag(t+\tau)\hat{a}_1(t+\tau)\hat{a}_2(t)\rangle
\nonumber\\
& +&\cos^4\xi\langle \hat{a}_1^\dag(t)\hat{a}_2^\dag(t+\tau)\hat{a}_2(t+\tau)\hat{a}_1(t)\rangle \nonumber \\
& -&\cos^2\xi \sin^2\xi\left(\langle
\hat{a}_1^\dag(t)\hat{a}_2^\dag(t+\tau)\hat{a}_1(t+\tau)\hat{a}_2(t)\rangle \right. \nonumber\\
& +&\left. \langle
\hat{a}_2^\dag(t)\hat{a}_1^\dag(t+\tau)\hat{a}_2(t+\tau)\hat{a}_1(t)\rangle
\right) \,. \label{g2_34}
\end{eqnarray}
In what follows we assume ideal mode-matched beams in inputs 1 and
2. Hence the bracket notation corresponds to time expectations
only.

In Eq.~(\ref{g2_34}), photon annihilation operators of channels 1
and 2 are due to the radiation field of a general single two-level
emitter. In the far field, this field annihilation operator is
given by the source-field relationship as
\begin{eqnarray}
\hat{a}(t)=A(\mathbf{r})
\hat{\sigma}_{ge}\left(t-\frac{\left|\mathbf{r}\right|}{c}\right)\,,\label{source_free}
\end{eqnarray}
where $A(\mathbf{r})$ is a time-independent proportionality
factor~\cite{Loudon}. This linear relationship allows for
substitution of photon annihilation and creation operators by
dipole projection operators $\hat{\sigma}_{ge}$ and
$\hat{\sigma}_{eg}$ respectively in Eq.~(\ref{g2_34}). Using the
assumption that both of the emitters are independent and have
identical expectation values and coherence functions, we arrive at
\begin{eqnarray}
G^{(2)}_{34}(t,\tau) &=& \left[ \left(\cos^4\xi+\sin^4\xi\right)
\langle \hat{\sigma}_{ee}(t)\rangle \langle
\hat{\sigma}_{ee}(t+\tau)\rangle \right. \nonumber \\&&
-2\cos^2\xi\sin^2\xi|\widetilde{G}^{(1)}(t,\tau)|^2] \,
\left|A(\mathbf{r})\right|^4 \,.\label{g2_34_sigma_simple}
\end{eqnarray}
In this equation $\widetilde{G}^{(1)}(t,\tau)$ represents the
unnormalized first-order coherence function
\begin{eqnarray}
\widetilde{G}^{(1)}(t,\tau)= \langle
\hat{\sigma}_{eg}(t+\tau)\hat{\sigma}_{ge}(t)\rangle\,.
\label{g1_interf}
\end{eqnarray}
For a balanced beam-splitter, $\theta=\pi/4$,
Eq.~(\ref{g2_34_sigma_simple}) simplifies to
\begin{eqnarray}
\widetilde{G}^{(2)}_{34}(t,\tau) & \equiv &
\frac{G^{(2)}_{34}(t,\tau)}{|A(\mathbf{r})|^4} \nonumber \\ & = &
\frac{1}{2} \left(\langle \hat{\sigma}_{ee}(t)\rangle \langle
\hat{\sigma}_{ee}(t+\tau)\rangle -
|\widetilde{G}^{(1)}(t,\tau)|^2\right)\,.
\label{g2_34_sigma_very_simple}
\end{eqnarray}
This is the expression of the unnormalized second order coherence
function in terms of the dipole projection operators that we will
use in the remaining of this section.

Under pulsed excitation further considerations need to be taken
into account to normalize this equation. Before this discussion
however, we note that under continuous wave excitation,
Eq.~(\ref{g2_nor}) reveals the normalized second order coherence
function
\begin{eqnarray}
g^{(2)}_{34}(t,\tau) &=& \frac{1}{2} \left(1-
\frac{\left|\widetilde{G}^{(1)}(t,\tau)\right|^2}{\langle
\hat{\sigma}_{ee}(t)\rangle_{ss}^2}\right)\nonumber\\
&=& \frac{1}{2} \left(1- \left|g^{(1)}(\tau)\right|^2\right)\,,
\label{normalized_g2_34_sigma_very_simple}
\end{eqnarray}
where $\langle \hat{\sigma}_{ee}(t)\rangle_{ss}$ represents the
steady state population density of the excited state.

\begin{figure}
\begin{center}
  \centerline{\psfig{figure=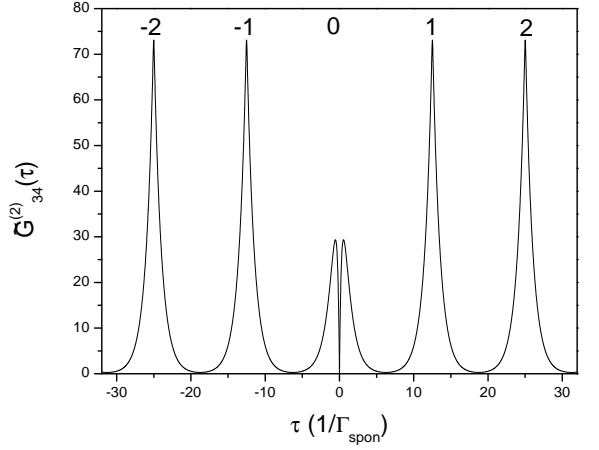,angle=0,width=3.0in}}
  \caption{Unnormalized coincidence detection rate, $\widetilde{G}^{(2)}_{34\_exp}$, of an incoherently pumped quantum dot.
  Parameter values are: $\Gamma_{relax}=100\Gamma_{spon}$, $\gamma_{deph}=\Gamma_{spon}$, each laser pulse is
  a Gaussian with pulsewidth $0.05/\Gamma_{spon}$ and peak Rabi frequency $35\Gamma_{spon}$.}
\goodvgap
\label{two_level_sim}
\end{center}
\end{figure}
Experimental determination of the cross-correlation function
relies on ensemble averaging coincidence detection events. Hanbury
Brown and Twiss setup is frequently used in these experiments
where the experimentally relevant cross-correlation function
\begin{eqnarray}
\widetilde{G}^{(2)}_{34\_exp}(\tau)=\lim_{T\rightarrow\infty}\int_{0}^{T}{\widetilde{G}^{(2)}_{34}(t,\tau)\,dt}
\,, \label{bs_photocount}
\end{eqnarray}
is measured. The total detection time $T$ is long compared to the
single photon pulsewidth ($T\rightarrow\infty$) in these
experiments.

In Fig.~\ref{two_level_sim} we plot an exemplary calculation of
$\widetilde{G}^{(2)}_{34\_exp}(\tau)$ for an incoherently pumped,
dephased quantum dot considering a series of 6 pulses. This
calculation is done by the integration of
$\widetilde{G}^{(2)}_{34}(t,\tau)$ (Eq.~(\ref{bs_photocount})),
while $\widetilde{G}^{(2)}_{34}(t,\tau)$ is calculated using the
optical Bloch equations and the quantum regression theorem. We
will detail these calculations in the following subsections. In
such calculations, the area of the peak around $\tau\sim0$
($0^{th}$ peak) gives the unnormalized coincidence detection
probability when two photons are incident in
different inputs of the beam splitter. This area should be
normalized by the area of the other peaks: Absence of two-photon
interference implies $0^{th}$ peak and other peaks to be
identical, whereas in total two-photon interference, $0^{th}$ peak
has vanishing area. This normalized measure of two-photon
interference is
\begin{eqnarray}
p_{34}&=&\frac{\int_{t=0}^\infty\int_{\tau,0}{\widetilde{G}^{(2)}_{34}(t,\tau)\,dt\,d\tau}}
{\int_{t=0}^\infty\int_{\tau,n}\widetilde{G}^{(2)}_{34}(t,\tau)
\,dt\,d\tau} \,.\label{norm_p34}
\end{eqnarray}
In the numerator, integral in $\tau$ is taken over the $0^{th}$
peak, whereas in the denominator this integral is taken over the
$n^{th}$ peak where $n=\pm 1,\pm 2,...$.

We now simplify Eq.~(\ref{norm_p34}) further using the periodicity
with respect to $\tau$ and $t$. First simplification is due to
periodicity in $\tau$ which is apparent in the periodicity of the
peaks other than $0^{th}$ peak in Fig.~\ref{two_level_sim}. The
area of these peaks is given by
\begin{eqnarray}
\int_0^\infty{\langle\hat{\sigma}_{ee}(t)\rangle \langle
\hat{\sigma}_{ee}(t+\tau-nT_{pulse})\rangle \,dt}\,,
\end{eqnarray}
for $n=\pm 1,\pm 2,...$ This is due to the vanishing
$\widetilde{G}^{(1)}(t,\tau)$ for absolute delay times larger than
single photon coherence time. Hence the normalized coincidence
probability can also be represented as
\begin{eqnarray}
p_{34}&=&\frac{\int_{t=0}^\infty\int_{\tau,0}{\widetilde{G}^{(2)}_{34}(t,\tau)\,dt\,d\tau}}
{\int_{t=0}^\infty\int_{\tau,0}\langle \hat{\sigma}_{ee}(t)\rangle
\langle \hat{\sigma}_{ee}(t+\tau)\rangle \,dt\,d\tau}\,.
\label{tau_simple}
\end{eqnarray}

Periodicity of $\widetilde{G}^{(2)}_{34}(t,\tau)$ and $\langle
\hat{\sigma}_{ee}(t)\rangle \langle
\hat{\sigma}_{ee}(t+\tau)\rangle$ in $t$ further simplifies
Eq.~(\ref{tau_simple}) to
\begin{eqnarray}
p_{34}&=&\frac{N\int_{t=0}^{T_{pulse}}\int_{\tau,0}{\widetilde{G}^{(2)}_{34}(t,\tau)\,dt\,d\tau}}
{N\int_{t=0}^{T_{pulse}}\int_{\tau,0}\langle
\hat{\sigma}_{ee}(t)\rangle \langle
\hat{\sigma}_{ee}(t+\tau)\rangle \,dt\,d\tau}\nonumber\\
&=&\frac{\int_{t=0}^{T_{pulse}}\int_{\tau,0}{\widetilde{G}^{(2)}_{34}(t,\tau)\,dt\,d\tau}}
{\int_{t=0}^{T_{pulse}}\int_{\tau,0}\langle
\hat{\sigma}_{ee}(t)\rangle \langle
\hat{\sigma}_{ee}(t+\tau)\rangle \,dt\,d\tau}
\,,\label{t_tau_simple}
\end{eqnarray}
where N represents the number of pulses considered in the
calculation.

Eq.~(\ref{t_tau_simple}) is the final result of the
simplifications and is used in the rest of this section. It is
important to note that this equation enables us to obtain the
normalized coincidence probability, $p_{34}$, by considering only
a single laser pulse. This greatly improves the efficiency of the
simulations.

There are two limitations of our method of calculation. Firstly,
the optical Bloch equation description does not take into account
laser broadening induced by amplitude or phase fluctuations.
Secondly, in the case of a quantum dot, an upper limit to laser
broadening may arise due to the biexciton splitting ($\sim3.5~meV$
at cryogenic temperatures) and Zeeman splitting ($\sim1~meV$ for
an applied field of 10~Tesla). Overall, these restrictions should
put a lower limit of $\sim1\times10^{-12}$ s to the laser
pulsewidth. This lower limit is always exceeded in our
calculations.

\subsection{Single photon source based on an incoherently pumped quantum dot}\label{3_level_sec}
Various demonstrations of single photon sources based on
solid-state emitters have been reported in recent years. Single
quantum
dots~\cite{michler2000,santori2001,zwiller2001,moreau2001,yuan2002},
single molecules~\cite{demartini1996,brunel1999,lounis2000}, and
single N vacancies~\cite{kurtsiefer2000,beveratos2001} were used
in these demonstrations where pulsed excitation of a high energy
state followed by a fast relaxation and excited state
recombination proved to be a very convenient method to generate
triggered single photons. This method of incoherent pumping
ensured the detection of at most one photon per pulse, provided
that the laser had sufficiently short pulses, and large pulse
separations.

\begin{figure}
\begin{center}
  \centerline{\psfig{figure=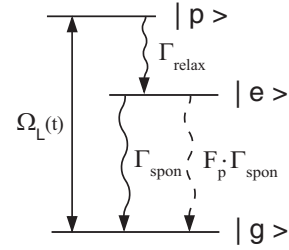,height=1.3in}}
  \caption{Model of an incoherently pumped single quantum dot. Dashed line
  demonstrates the generated single photons via cavity leakage.}
\goodvgap
\label{undephased_QD}
\end{center}
\end{figure}

In the following, we extensively consider the case of quantum dots
and analyze two-photon interference among photons emitted from an
incoherently pumped quantum dot. In such a three-level scheme
(Fig.~\ref{undephased_QD}), time-jitter induced by the fast
relaxation ($\Gamma_{relax}$) and dephasing in
$|e\rangle$-$|g\rangle$ transition are the sources of non-ideal
two-photon interference. We investigate these effects first under
continuous wave, then under pulsed excitation.

\subsubsection{Continuous wave excitation}
Under continuous wave excitation, $\widetilde{G}^{(1)}(t,\tau)$ is
calculated by applying quantum regression theorem~\cite{Loudon} to
the optical Bloch equation for $\langle \hat{\sigma}_{eg}(t)
\rangle$, revealing
\begin{eqnarray}
\frac{d\widetilde{G}^{(1)}(t,\tau)}{d\tau}=-\gamma
\widetilde{G}^{(1)}(t,\tau) \,,\label{bloch_cw}
\end{eqnarray}
where $\gamma=\frac{\Gamma_{spon}}{2}+\gamma_{deph}$ is the total
coherence decay rate of $|e\rangle$-$|g\rangle$ transition. Here
$\gamma_{deph}$ denotes dephasing caused by all reservoirs other
than that of the radiation field.

Following the solution of Eq.~(\ref{bloch_cw}), using the initial
condition $\widetilde{G}^{(1)}(t,0)=\langle \hat{\sigma}_{ee}(t)
\rangle_{ss}$, the normalized coincidence detection probability is
obtained by Eq.~(\ref{normalized_g2_34_sigma_very_simple}) as
\begin{eqnarray}
g^{(2)}_{34}(\tau) &=& \frac{1}{2} \left(1-e^{-2\gamma\tau}\right)
\,.\label{cw_QD_sigma_very_simple}
\end{eqnarray}
Hence, for the continuous wave excitation case,
indistinguishability is solely determined by the total coherence
decay rate in $|e\rangle$-$|g\rangle$ transition. Decay time of
the normalized coincidence detection probability is $1/2\gamma$.

\subsubsection{Pulsed excitation}\label{incoh_pulsed}
A more detailed study of Bloch equations is necessary for the case
of pulsed excitation. The interaction Hamiltonian of the system
depicted in Fig.~\ref{undephased_QD} is
\begin{eqnarray}
\hat{H}_{int} = i\hbar \Omega_L ( \hat{\sigma}_{pg} -
\hat{\sigma}_{gp})\,.
\end{eqnarray}
The master equation
\begin{eqnarray}
\frac{d}{dt} \hat{\rho} &=& \frac{1}{i\hbar}
\left[\hat{H}_{int},\hat{\rho}\right]+\frac{\Gamma_{relax}}{2}\left(2\hat{\sigma}_{gp}
\hat{\rho} \hat{\sigma}_{pg} - \hat{\sigma}_{pp} \hat{\rho} -
\hat{\rho} \hat{\sigma}_{pp} \right) \nonumber\\
&&+\frac{\Gamma_{spon}}{2}\left(2\hat{\sigma}_{ge} \hat{\rho}
\hat{\sigma}_{eg} - \hat{\sigma}_{ee} \hat{\rho} - \hat{\rho}
\hat{\sigma}_{ee}\right)\,,
\end{eqnarray}
is used to derive the optical Bloch equations. As described
previously, calculation of $p_{34}$ follows the solution of the
optical Bloch equations and Eq.~(\ref{bloch_cw}) considering a
single laser pulse.

\begin{figure}
\begin{center}
  \centerline{\psfig{figure=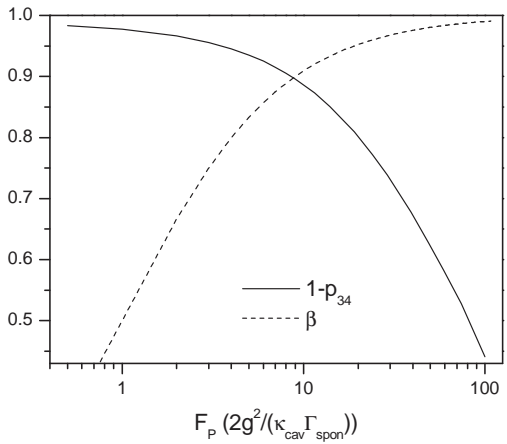,angle=0,height=2.3in}}
\caption{Dependence of indistinguishability and collection
efficiency on the cavity-induced decay rate
($(F_P+1)\Gamma_{spon}$) of a quantum dot. Parameter values are:
$\Gamma_{spon}=10^{9}~\textrm{s}^{-1}$,
$\Gamma_{relax}=10^{11}~\textrm{s}^{-1}$, $\gamma_{deph}=0$,
excitation laser is a Gaussian beam with a pulsewidth of
$10^{-11}$~s. Peak laser Rabi frequency is changed between
$1.1\times10^{11}$ and $0.93\times10^{11}~\textrm{s}^{-1}$.}
  \label{jitter_sim}
\end{center}
\end{figure}
We now study the dependence of indistinguishability, ($1-p_{34}$),
on the cavity-induced decay rate ($(F_P+1)\Gamma_{spon}$) and
dephasing. In Fig.~\ref{jitter_sim}, we plot the collection
efficiency and indistinguishability as a function of the Purcell
factor, $F_P$, for a quantum dot with $\gamma_{deph}=0$. We assume
$\Gamma_{spon}=10^9~\textrm{s}^{-1}$ and
$\Gamma_{relax}=10^{11}~\textrm{s}^{-1}$. Peak laser Rabi
frequency is changed between $1.1\times10^{11}$ and
$0.93\times10^{11}~\textrm{s}^{-1}$ in order to achieve
$\pi$-pulse excitation for different Purcell factors. Collection
efficiency is calculated by $\beta=F_P/(F_P+1)$, assuming that
photons emitted to the cavity mode are collected with 100$\%$
efficiency. This assumption clearly constitutes an upper limit for
the actual collection efficiency for typical
microcavities~\cite{gerard2002}.

Fig.~\ref{jitter_sim} depicts one of the main results we present
in this paper. Due to the time-jitter induced by the relaxation
from the third level, there is a trade-off between collection
efficiency and indistuingishability. For a Purcell factor of 100
we calculate a maximum indistuingishability of 44~$\%$ with a
collection efficiency of 99~$\%$.

\begin{figure}
\begin{center}
  \centerline{\psfig{figure=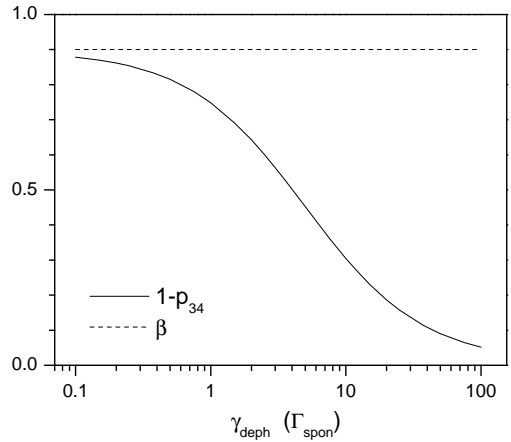,angle=0,height=2.3in}}
  \caption{Dependence of indistinguishability and collection efficiency on dephasing ($\gamma_{deph}$).
  $\Gamma_{spon}=10^{9}~\textrm{s}^{-1}$, $F_P=9$,
$\Gamma_{relax}=10^{11}~\textrm{s}^{-1}$, excitation laser is a
Gaussian beam with a pulsewidth of $10^{-11}$~s. Peak laser Rabi
frequency is $1.03\times10^{11}~\textrm{s}^{-1}$ achieving
$\pi$-pulse excitation.}
\label{dephasing_sim}
\end{center}
\end{figure}
The dependence of indistinguishability on dephasing is depicted in
Fig.~\ref{dephasing_sim}. As expected, dephasing has no effect on
the collection efficiency. On the other hand, indistinguishability
vanishes for $\gamma_{deph}>\Gamma_{spon}$. Since dephasing is
effectively a non-referred quantum state measurement, it results
in additional jitter in photon emission time.

To understand this effect, we should recall that dephasing of an
optical transition is equivalent to a non-referred quantum state
measurement that projects the emitter into either its excited or
ground state. Reciprocal dephasing rate $\gamma_{deph}^{-1}$ then
gives the average time interval between these state projections.
In this case, photon emission is restricted to take place in
between two subsequent measurement events, first (second) of which
projects the emitter into the excited (ground) state. While the
bandwidth of the emitted photon is then necessarily given by
$\gamma_{deph}$ due to energy-time uncertainty, its emission (i.e.
arrival) time will be randomly distributed within
$\Gamma_{spon}^{-1}$. Since the information about the random
emission times of any two photons is carried by the reservoir that
causes the dephasing process, the photons will no longer be
completely indistinguishable.

\subsection{Quantum dot single photon source based on a cavity-assisted spin-flip Raman transition}
Raman transition in a single three-level system strongly coupled
to a high-Q cavity provides an alternative single photon
generation scheme~\cite{C.K.Law:PRL96,C.K.Law:JMO97,A.Kuhn:APB99}.
In contrast to the incoherently pumped source discussed in
subsection~\ref{3_level_sec}, this scheme realizes a coherently
pumped single photon source that does not involve coupling to
reservoirs other than the one into which single photons are
emitted. It allows for pulse-shaping, and is suitable for quantum
state transfer~\cite{cirac1997}. In this part we discuss the
application of this scheme to quantum dots, and demonstrate that
arbitrarily high collection efficiency and indistinguishability
can simultaneously be achieved in this scheme.

\begin{figure}
\begin{center}
  \centerline{\psfig{figure=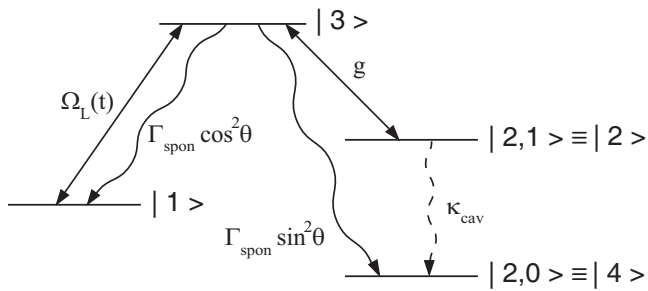,height=1.5in}}
\caption{Single photon source based on cavity-assisted spin-flip
Raman transition in a single quantum dot. Dashed line demonstrates
the generated single photons via cavity leakage.}
  \goodvgap
  \label{raman_sch}
\end{center}
\end{figure}

A quantum dot doped with a single conduction-band electron
constitutes a three-level system in the $\Lambda$-configuration
under constant magnetic fields along x-direction
(Fig.~\ref{raman_sch})~\cite{imamoglu99}. Lowest energy conduction
and valence band states of such a quantum dot are represented by
$|m_x=\pm1/2\rangle$ and $|m_z=\pm3/2\rangle$ respectively due to
the strong z-axis confinement, typical of quantum dots. The
magnetic field results in the Zeeman splitting of the spin up
($|m_x=1/2\rangle$) and down ($|m_x=-1/2\rangle$) levels in the
conduction band. Considering an electron g-factor of 2 and an
applied field of 10~Tesla which is available from typical
magneto-optical cryostats, the splitting is expected to be
$\sim$1~meV. At cryogenic temperatures, this splitting is much
larger than other broadenings in consideration, thus a three-level
system in the $\Lambda$-configuration is obtained. We emphasize
that none of the experimental measurements carried out on
self-assembled quantum dots yield any signatures of Auger
recombination processes for trion (2 electron and one hole system)
or biexciton transitions. In particular, lifetime measurements
carried out on biexcitons gave $\tau_{biexc}\sim\tau_{exc}/1.5$,
indicating the absence of Auger enhancement of biexciton
decay~\cite{Kiraz2002}.

We assume that an x-polarized laser pulse is applied resonantly
between levels $|m_x=1/2\rangle$ and $|m_z=3/2\rangle$ (or
$|m_z=-3/2\rangle$) while levels $|m_x=-1/2\rangle$ and
$|m_z=3/2\rangle$ (or $|m_z=-3/2\rangle$) are strongly coupled via
a resonant y-polarized cavity mode. Considering the number of
cavity photons to be limited to 0 and 1, the electronic energy
level $|m_x=-1/2\rangle$, can be represented by the levels
$|m_x=-1/2,1\rangle$ and $|m_x=-1/2,0\rangle$ corresponding to 1
and 0 cavity photon respectively. We will abbreviate the energy
levels $|m_x=1/2\rangle$, $|m_z=3/2\rangle$, $|m_x=-1/2,1\rangle$,
and $|m_x=-1/2,0\rangle$ as $|1\rangle$, $|2\rangle$, $|3\rangle$,
and $|4\rangle$ respectively.

In such a three-level system, Raman transition induced by the
laser and cavity fields together with the finite cavity leakage
rate, $\kappa_{cav}$, enable the generation of a single cavity
photon per pulse. For large field couplings, level $|3\rangle$ can
be totally bypassed resulting in ideal coherent population
transfer between levels $|1\rangle$ and $|2\rangle$. This single
photon source has therefore the potential to achieve 100$\%$
collection efficiency together with ideal two-photon interference.
This scheme is to a large extent insensitive to quantum dot size
fluctuations and may enable the use of different quantum dots in
simultaneous generation of indistinguishable photons, provided
that the cavity resonances and the electron g-factors are
identical. Variations in the electron g-factor between different
quantum dots would limit the photon indistinguishability due to
spectral mismatch between the generated photons: We do not
consider this potential limitation in this paper. In general,
spontaneous emission and dephasing in $|3\rangle$-$|2\rangle$
transition are the principal sources of non-ideal two-photon
interference and decreased collection efficiency in this scheme.
The ultimate limit for photon indistinguishability due to jitter
in emission time is given by spin decoherence of the ground state.

Such a single photon source has been recently demonstrated using
single cold atoms trapped in a high-Q Fabry-Perot
cavity~\cite{A.Kuhn:PRL02}. Due to the limited trapping times, at
most only 7 photons were emitted by a single atom in this
demonstration. Practical realizations of this scheme also require
a means to bring the system from level $|4\rangle$ to $|1\rangle$
at the end of each single photon generation event. In
Ref.~\cite{A.Kuhn:PRL02} this was achieved by a recycling laser
pulse. The applied recycling laser pulse determines the end of the
single-photon pulse and can in principle limit the collection
efficiency for systems with long spontaneous emission lifetimes.
In the case of quantum dots, recycling can be achieved by a
similar laser pulse applied between levels $|4\rangle$ and
$|3\rangle$. An alternative recycling mechanism can be the
application of a Raman $\pi$-pulse, generated by two detuned laser
pulses satisfying the Raman resonance condition between levels
$|4\rangle$ and $|1\rangle$.

We now discuss the numerical analysis of this system which is
described by the interaction Hamiltonian
\begin{eqnarray}
\hat{H}_{int} = i\hbar g ( \hat{\sigma}_{32} - \hat{\sigma}_{23})
+ i\hbar \Omega_L ( \hat{\sigma}_{31} - \hat{\sigma}_{13})\,.
\end{eqnarray}
We use the master equation
\begin{eqnarray}
\frac{d}{dt} \hat{\rho} &=& \frac{1}{i\hbar}
\left[\hat{H}_{int},\hat{\rho}\right]+\kappa_{cav}\left(2\hat{\sigma}_{42}
\hat{\rho} \hat{\sigma}_{24} - \hat{\sigma}_{22} \hat{\rho} -
\hat{\rho} \hat{\sigma}_{22} \right) \nonumber\\
&&+\frac{\Gamma_{spon}\cos^2\theta}{2}\left(2\hat{\sigma}_{13}
\hat{\rho} \hat{\sigma}_{31} - \hat{\sigma}_{33} \hat{\rho} -
\hat{\rho} \hat{\sigma}_{33} \right) \nonumber\\
&&+\frac{\Gamma_{spon}\sin^2\theta}{2}\left(2\hat{\sigma}_{43}
\hat{\rho} \hat{\sigma}_{34} - \hat{\sigma}_{33} \hat{\rho} -
\hat{\rho} \hat{\sigma}_{33} \right)\,,
\end{eqnarray}
to derive the optical Bloch equations. In the presence of
dephasing caused by reservoirs other than the radiation field ($\gamma_{deph}$), we define the total coherence decay rate in transitions from level $|3\rangle$ as $\gamma=\frac{\Gamma_{spon}}{2}+\gamma_{deph}$.
Branching of spontaneous emission from level $|3\rangle$ to levels
$|1\rangle$ and $|4\rangle$ is indicated by $\cos^2\theta$ and
$\sin^2\theta$, respectively, as shown in Fig.~\ref{raman_sch}.

$\widetilde{G}^{(1)}(t,\tau)=\langle
\hat{\sigma}_{24}(t+\tau)\hat{\sigma}_{42}(t)\rangle$ is
calculated by applying the quantum regression theorem to the
optical Bloch equations for $\hat{\sigma}_{14}$,
$\hat{\sigma}_{24}$, and $\hat{\sigma}_{34}$. The following set of
differential equations are then obtained
\begin{eqnarray}
\frac{d}{d\tau}F(t,\tau)&=&-\Omega_L(t)H(t,\tau)\,,\nonumber\\
\frac{d}{d\tau}\widetilde{G}^{(1)}(t,\tau)&=&-gH(t,\tau)-\kappa_{cav}\widetilde{G}^{(1)}(t,\tau)\,,\label{raman_qr}\\
\frac{d}{d\tau}H(t,\tau)&=&\Omega_L(t)F(t,\tau)+g\widetilde{G}^{(1)}(t,\tau)-\gamma
H(t,\tau)\,.\nonumber
\end{eqnarray}
The variables: $\widetilde{G}^{(1)}(t,\tau)=\langle
\hat{\sigma}_{24}(t+\tau)\hat{\sigma}_{42}(t)\rangle$,
$F(t,\tau)=\langle
\hat{\sigma}_{14}(t+\tau)\hat{\sigma}_{42}(t)\rangle$, and
$H(t,\tau)=\langle
\hat{\sigma}_{34}(t+\tau)\hat{\sigma}_{42}(t)\rangle$ have initial
conditions $\widetilde{G}^{(1)}(t,0)=\langle
\hat{\sigma}_{22}(t)\rangle$, $F(t,0)=\langle
\hat{\sigma}_{12}(t)\rangle$, and $H(t,0)=\langle
\hat{\sigma}_{32}(t)\rangle$.

Following the solutions of the optical Bloch equations and the set
of Eqs.~(\ref{raman_qr}), normalized coincidence detection
probability, $p_{34}$, is calculated using
Eq.~(\ref{t_tau_simple}) as described in
section~\ref{sec_normalization}. Assuming ideal detection of the
photons emitted to the cavity mode, we calculate the collection
efficiency by the number of photons emitted from the cavity
\begin{eqnarray}
n=2\kappa_{cav}\int_0^\infty\langle
\hat{\sigma}_{22}(t)\rangle\,dt\,.
\end{eqnarray}

\begin{figure}
\begin{center}
  \centerline{\psfig{figure=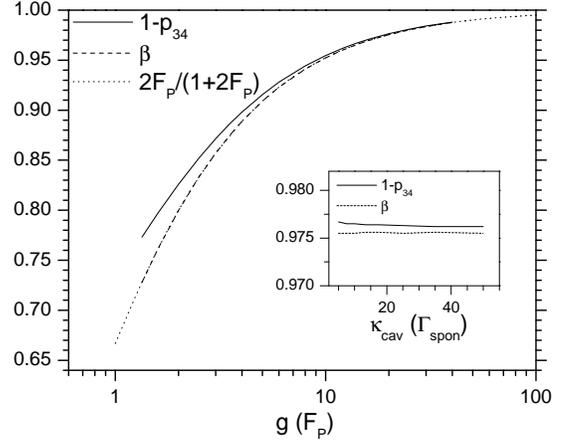,angle=0,height=2.3in}}
  \caption{Dependence of indistinguishability and collection efficiency
  on cavity coupling. Parameter values are: $\Gamma_{spon}=1$, $\kappa_{cav}=10$,
  $\gamma_{deph}=0$, $\theta=\pi/4$, a Gaussian pulse with pulsewidth=10,
  peak laser Rabi frequency is changed between 0.75 and 2.8 . Inset: Dependence of indistinguishability
  and collection efficiency on $\kappa_{cav}$ for a constant $F_P$ of 20.
  Parameter values are: $\Gamma_{spon}=1$, $\gamma_{deph}=0$, $\theta=\pi/4$, laser
  pulsewidth of $10/\Gamma_{spon}$, peak laser Rabi frequency of 1.9 - 2.1~.}
\label{raman_sim}
\end{center}
\end{figure}

Our principal numerical results are depicted in
Fig.~\ref{raman_sim} where we consider a dephasing-free system,
and analyze the dependence of the collection efficiency and
indistinguishability on the cavity coupling. In these calculations
we assume a potential quantum dot cavity-QED system with
relatively small cavity decay rate of
$\kappa_{cav}=10\Gamma_{spon}$~\cite{QDcavity}. Laser pulse is
chosen to be Gaussian with a constant pulsewidth. The peak laser
Rabi frequency is increased with increased cavity coupling in
order to reach the onset of saturation in the emitted number of
photons. The large pulsewidth of 10 ensures the operation in the
regime where collection efficiency and indistinguishability are
independent of the pulsewidth. All other parameters are kept
constant at their values noted in the figure caption. We choose
both spontaneous emission channels to be equally present
($\theta=\pi/4$).

In contrast to the incoherently pumped single photon source,
Fig.~\ref{raman_sim} shows that arbitrarily high
indistinguishability and collection efficiency can simultaneously
be achieved with better cavity coupling using this scheme. For a
cavity coupling that corresponds to a Purcell factor of 40
($F_P=2g^2/(\kappa_{cav}\Gamma_{spon})=40$), our calculations
reveal 99~$\%$ indistinguishability together with 99~$\%$
collection efficiency. This regime of operation is readily
available in current state-of-the-art experiments with
atoms~\cite{hood2000}. While such a Purcell factor has not been
observed for solid-state emitters in microcavity structures to
date, recent theoretical~\cite{vuckovic2003} and
experimental~\cite{QDcavity,akahane2003} progress indicate that the
aforementioned values could be well within reach.

As expected, the dependence of $\beta$ on cavity coupling is
exactly given by $2F_P/(1+2F_P)$. This is due to the spontaneous
emission from level $|3\rangle$ to $|4\rangle$, namely
$\Gamma_{spon}\sin^2\theta=\Gamma_{spon}/2$, which defines the
relevant Purcell factor. As shown in the inset in
Fig.~\ref{raman_sim}, our calculations considering different
$\kappa_{cav}$ values for a constant Purcell factor revealed
similar collection efficiency and indistinguishablity values.
Hence Purcell factor is the most important parameter in
determining the characteristics of this single photon source.

\begin{figure}
\begin{center}
  \centerline{\psfig{figure=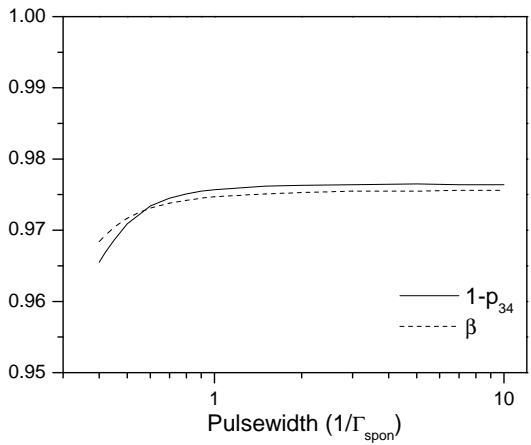,angle=0,height=2.3in}}
  \caption{Dependence of indistinguishability and collection efficiency on the Gaussian laser pulsewidth.
  Parameter values are: $\Gamma_{spon}=1$,
  $g=10$, $\kappa_{cav}=10$ ($F_P=20$), $\gamma_{deph}=0$, $\theta=\pi/4$.
Peak laser Rabi frequency is changed between 2.1 and 10.5 .}
\label{raman_pulsewidth}
\end{center}
\end{figure}

Achieving the regime of large indistinguishability and collection
efficiency together with small laser pulsewidths is also highly
desirable for efficient quantum information processing
applications. In this single photon source that relies on
cavity-assisted Raman transition, lower limits for the laser
pulsewidth are in general given by the inverse cavity coupling
constant ($g^{-1}$) and cavity decay rate
($\kappa_{cav}^{-1}$)~\cite{A.Kuhn:APB99}. We analyze the effect
of the laser pulsewidth to indistinguishability and collection
efficiency in Fig.~\ref{raman_pulsewidth}. In this figure we
consider the potential quantum dot cavity-QED system analyzed in
Fig.~\ref{raman_sim} ($\kappa_{cav}=10\Gamma_{spon}$) while
assuming a Purcell factor of 20 ($g=10\Gamma_{spon}$). As in the
previous cases, we change the maximum laser Rabi frequency for
different pulsewidth values in order to reach the onset of
saturation. For this system, we conclude that a minimum pulsewidth
of $1/\Gamma_{spon}$ is sufficient to achieve maximum
indistinguishability and collection efficiency.

\begin{figure}
\begin{center}
  \centerline{\psfig{figure=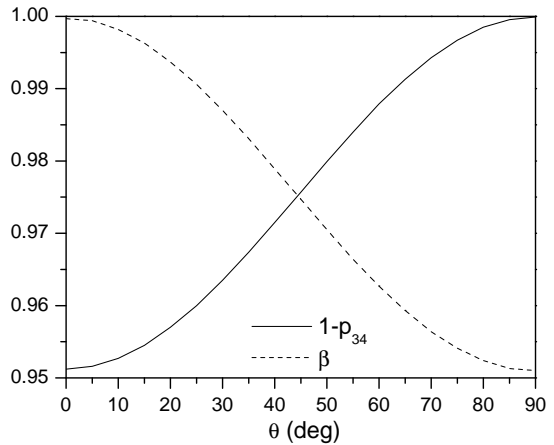,angle=0,height=2.3in}}
  \caption{Dependence of indistinguishability and collection efficiency on $\theta$.
  Parameter values are: $\Gamma_{spon}=1$,
  $g=10$, $\kappa_{cav}=10$ ($F_P=20$), $\gamma_{deph}=0$, $\theta=\pi/4$. A
Gaussian pulse is assumed with pulsewidth=1 and peak Rabi
frequency of 6.2 .}
\label{raman_theta}
\end{center}
\end{figure}

The two spontaneous emission channels from level $|3\rangle$ have
complementary effects on collection efficiency and
indistinguishability. Spontaneous emission from level $|3\rangle$
to $|1\rangle$ reduces indistinguishability while having no effect
on collection efficiency. This spontaneous emission channel,
$\Gamma_{spon}\cos^2\theta$, effectively represents a time-jitter
mechanism for single photon generation. In contrast, spontaneous
emission from level $|3\rangle$ to level $|4\rangle$ has no effect
on indistinguishability while reducing collection efficiency.
These effects are clearly demonstrated in Fig.~\ref{raman_theta}
where we plot the dependence of collection efficiency and
indistinguishability on $\theta$.

\begin{figure}
\begin{center}
  \centerline{\psfig{figure=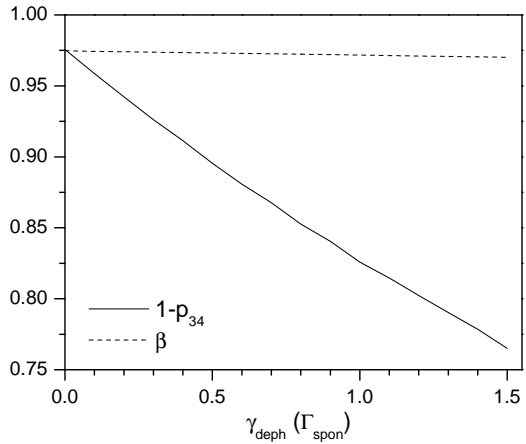,angle=0,height=2.3in}}
  \caption{Dependence of indistinguishability and collection efficiency on the dephasing rate.
  Parameter values
are: $\Gamma_{spon}=1$, $g=10$, $\kappa_{cav}=10$ ($F_P=20$),
$\theta=\pi/4$, a Gaussian laser pulse is assumed with
pulsewidth=1 and peak Rabi frequency of 6.2 .}
\label{raman_dephasing}
\end{center}
\end{figure}

Finally in Fig.~\ref{raman_dephasing} we analyze the dependence of
indistiguishability and collection efficiency on dephasing of
transitions from level $|3\rangle$. In contrast to the case of an
incoherently pumped quantum dot (Fig.~\ref{dephasing_sim}), there
is a small but non-zero dependence of collection effciency on
dephasing. For the parameters we chose, collection efficiencies of
0.975 and 0.970 were calculated for dephasing rates of $0$ and
$1.5\Gamma_{spon}$ respectively.

\section{Indistinguishability and nondeterministic linear-optics gates}\label{LOQC_sec}

Having determined the limits and dependence of photon collection
efficiency and indistinguishability on system configuration and
cavity parameters, we turn to the issue of photon
distinguishability effects on the performance of LOQC gates.
Related question of dependence on photon loss
\cite{KLM01,Ralph2003} and detection inefficiency
\cite{Glancy2002} have previously been analyzed. For semiconductor
single photon sources, photon loss can be minimized by increasing
collection efficiency, in principle, to near unity value.
Therefore, close to ideal photon emission can be achieved with
better cavity designs and coupling. However, as we have shown in
previous sections, an incoherently pumped semiconductor photon
source suffers heavily from emission time-jitter, especially for
large values of Purcell factor, while a semiconductor system based
on cavity-assisted spin-flip Raman transition shows promise for
near unity collection efficiency and indistinguishability. To
assess the cavity requirements for the latter system, we analyze
the reduction in gate fidelity arising from photon emission
time-jitter in a linear optics controlled phase gate, a key
element for most quantum gate constructions.

\begin{figure}
\begin{center}
  \centerline{\psfig{figure=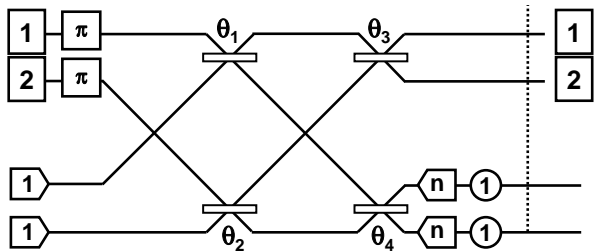,angle=0,height=1.3in}}
  \caption{Optical network realizing CS$_{180^\circ}$.}
\label{gate_setup}
\end{center}
\end{figure}

This non-deterministic gate operates as follows: Given a two-mode
input state of the form

\begin{eqnarray}
|\Psi_{\rm
in}\rangle=\bigl[\alpha|00\rangle+\beta|01\rangle+\delta|10\rangle+\gamma|11\rangle
\bigr]\,, \label{gate_in}
\end{eqnarray}

\noindent where $|\alpha|^2+|\beta|^2+|\gamma|^2+|\delta|^2=1$,
the state at the two output modes transforms into

\begin{eqnarray}
|\Psi_{\rm out}\rangle=
\bigl[\alpha|00\rangle+\beta|01\rangle+\delta|10\rangle+e^{\rm{i}\Phi}\gamma|11\rangle\
\bigr]\,, \label{gate_out}
\end{eqnarray}

\noindent with a certain probability of success, $|p|^2$.  A
realization of such a gate using all linear optical elements, two
helper single photons on demand, and two photon-number resolving
single-photon detectors is depicted in Fig.~\ref{gate_setup} for
the special case of $\Phi = \pi$~\cite{Knillgate,Knillgate2}. This
realization consists of two input modes for the incoming quantum
state to be transformed and two ancilla modes with a single helper
photon in each mode. After four beam splitters with settings
$\theta_{1}=\theta_{2}=-\theta_{3}=54.74^{\circ}$ and
$\theta_{4}=17.63^{\circ}$, postselection is performed via
photon-number measurements on output modes 3 and 4. Conditional to
single-photon detection in each of these modes, the quantum state
in Eq.~(\ref{gate_in}) is transformed into Eq. (\ref{gate_out}).
The probability of success for this construction is $2/27$, which
is slightly better then $1/16$, the probability of success of the
original proposal using only one helper photon with two ancilla
modes~\cite{KLM01}.

This is the probability of success for ideal systems comprising
indistinguishable photons, and unity efficiency number resolving
single photon detectors. We now proceed to investigate the effects
of photon distinguishability arising from physical constraints of
the single photon sources in consideration. In the presence of a
temporal jitter, $\epsilon$, in the photon emission time, a single
photon state can be represented as

\begin{eqnarray}
|1\rangle=\int \rm{d}\omega f(\omega)e^{\rm{i}\omega
\epsilon}a_{j}^{\dagger}(\omega)|0\rangle \,, \label{gate_photon}
\end{eqnarray}

\noindent where $f(\omega)$ is the spectrum of the photon
wave-packet. For photons from a quantum dot in a cavity, the
function $f(\omega)$ is a Lorentzian yielding a double-sided
exponential dip in the Hong-Ou-Mandel
interference~\cite{hong1987}. In the presence of relative time
jitter, the visibility of interference is obtained after ensemble
averaging over the time-jitter $\epsilon$ in the range $[0,
\epsilon_{0}]$ yielding the relation

\begin{eqnarray}
V(\epsilon_{0})=\frac{1}{\epsilon_{0}/\tau}(1-\text{e}^{-\epsilon_{0}
/ \tau}) \,, \label{visibility_func}
\end{eqnarray}

\noindent for a uniform distribution. In order to analyze
time-jitter effects on the fidelity of the quantum gate shown in
Fig.~\ref{gate_setup}, we introduce a time-jitter for the helper
photon in mode 4. For clarity, we keep the remaining photons in
other modes ideal and indistinguishable. The symmetry of the gate
ensures that each introduced time-jitter adds to the power
dependence of the overall error.

Rewriting Eq. (\ref{gate_photon}) as

\begin{eqnarray}
|1\rangle=\int \rm{d}\omega f(\omega)[1-(1-e^{\rm{i}\omega
\epsilon})] c_{j}^{\dagger}(\omega)|0\rangle \,,
\label{gate_photon2}
\end{eqnarray}

\noindent allows us to represent the output state in terms of the
ideal output state and the time-jitter dependent part
$|\Phi(\epsilon)\rangle$:

\begin{eqnarray}
|\overline{\Psi}_{\rm out}\rangle=|\Psi_{\rm
out}\rangle-|\Phi(\epsilon)\rangle \,. \label{gate_modified}
\end{eqnarray}

\noindent Using the definition of the gate fidelity for a
particular $|\overline{\Psi}_{\rm out}\rangle$

\begin{eqnarray}
F_{|\overline{\Psi}_{\rm out}\rangle} = \frac{|\langle \Psi_{\rm
out}|\overline{\Psi}_{\rm out}\rangle|^2}{\langle \Psi_{\rm
out}|\Psi_{\rm out}\rangle \langle \overline{\Psi}_{\rm out}|
\overline{\Psi}_{\rm out}\rangle} \,, \label{gate_fidelity}
\end{eqnarray}

\noindent with Eq. \ref{gate_modified} we obtain

\begin{eqnarray}
F_{|\overline{\Psi}_{\rm out}\rangle} =
\frac{|p|^2-2\rm{Re}\left[{\langle \Psi_{\rm
out}|\Phi(\epsilon)\rangle}\right]+\frac{|\langle \Psi_{\rm
out}|\Phi(\epsilon)\rangle|^2}{|p|^2}}{|p|^2-2\rm{Re}\left[{\langle
\Psi_{\rm out}|\Phi(\epsilon)\rangle}\right]+\langle
\Phi(\epsilon)|\Phi(\epsilon)\rangle}
 \,,
\label{gate_fidelity1.5}
\end{eqnarray}

\noindent where $|p|^2=\langle \Psi_{\rm out}|\Psi_{\rm
out}\rangle$. Given the particular realization of this gate as
depicted in Fig.~\ref{gate_setup}, the overall gate fidelity takes
the form

\begin{eqnarray}
F=\text{min}\left[ \bigl<F_{|\overline{\Psi}_{\rm
out}\rangle}\bigr>_{\epsilon}^{\epsilon_{0}}\right] =
\frac{c_{0}+c_{1}V(\epsilon_{0})+c_{2}V^2(\epsilon_{0})}{d_{0}+d_{1}V(\epsilon_{0})+d_{2}V^2(\epsilon_{0})}
 \,,
\label{gate_fidelity2}
\end{eqnarray}

\noindent where $\langle\cdot\rangle_{\epsilon}^{\epsilon_{0}}$
denotes ensemble averaging over time-jitter $\epsilon$ in the
range $[0,\, \epsilon_{0}]$ using an appropriate weight function
and $V(\epsilon_{0})$ is the degree of indistinguishability, or
the corresponding visibility in a Hong-Ou-Mandel interference. The
coefficients $c_{i}$ and $d_{i}$ in Eq. (\ref{gate_fidelity2})
depend not only on the gate properties such as the probability of
success, but also on the initial input state through the
coefficients $\alpha$, $\beta$, and $\gamma$. Consequently, the
gate fidelity becomes a function of the properties of the initial
input state.

\begin{figure}
\begin{center}
  \centerline{\psfig{figure=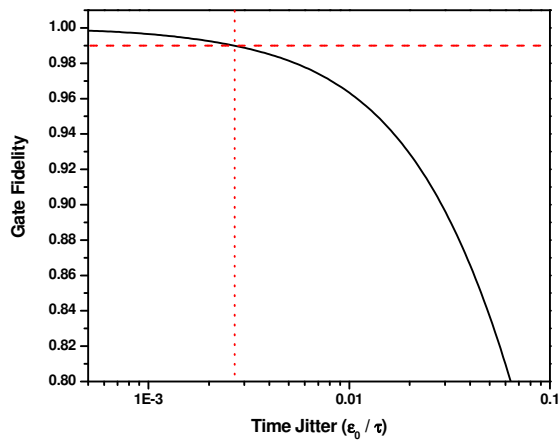,angle=-90,height=2.3in}}
  \caption{Dependence of nonlinear sign gate fidelity $F$, on normalized time-jitter $\epsilon$.
  The horizontal line indicates the 99\% fidelity threshold. The
vertical line indicates tolerable time-jitter threshold.}
\label{gate_Fplot}
\end{center}
\end{figure}

A plot for minimum gate fidelity (corresponding to a $|11\rangle$
input state) found after extensive search over a set of initial
input states is shown in Fig.~\ref{gate_Fplot} as a function of
time-jitter normalized to photon pulsewidth $(\epsilon_{0}/\tau)$.
As is evident from the graph, time-jitter on the order of 0.3\% is
the limiting case in order to achieve fidelity of 99\%. For an
incoherently pumped quantum dot single photon source as analyzed
in section~\ref{3_level_sec}, the emission time-jitter is on the
order of $1\times10^{-11}$ s. Thus, for single photon pulsewidth
on the order of $1\times10^{-9}$ s, this fidelity threshold cannot
be satisfied. As is also clear from Fig.~\ref{jitter_sim}, this
corresponds to a Purcell factor of order unity and collection
efficiency of about 50\%. In a cavity-assisted spin-flip Raman
transition, however, indistinguishability and collection
efficiency are both shown to increase in Fig.~\ref{raman_sim} as
the Purcell factor increases. This, in turn, casts a single
constraint on the cavity quality factor, requiring $F_{P}\ge40$,
in order to achieve both indistinguishability and collection
efficiency required for gate operations for LOQC. This threshold
for cavity quality factor is within the realistic values to date.
We emphasize that so far there has been no calculation on the
maximum allowed time-jitter error for LOQC scheme~\cite{KLM01}.

\section{Conclusions}

We analyzed the effects of cavity coupling, spontaneous emission
rate, dephasing, and laser pulsewidth on indistinguishability and
collection efficiency for two distinct types of single photon
sources based on two and three-level emitters. We showed that, in
contrast to incoherently pumped systems, a single photon source
based on cavity-assisted spin-flip Raman transition has the
potential to simultaneously achieve high levels of
indistinguishability and collection efficiency. For this system,
in the absence of dephasing, 99~$\%$ indistinguishability and
collection efficiency are achieved for a Purcell factor of 40. Our
analysis revealed that strong coupling regime of cavity-QED
($g>\{\gamma,\kappa_{cav}\}$) is not a requirement for optimum
operation while, in the presence of dephasing, the characteristics
of the system is determined by $g^2/\kappa_{cav}\gamma_{deph}$
rather than the Purcell factor. The desired regime of operation,
i.e. Purcell factor of 40 in the absence of dephasing, is readily
available for atoms in high-Q Fabry-Perot cavities. It is also
within the reach for solid-state based single photon sources
embedded in microcavity structures given current technology. We
also analyzed the reduction in gate fidelity arising from photon
emission-time-jitter in a linear optics controlled sign gate. We
found that the aforementioned Purcell regime provides gate
performance with error $<\,1\%$ using the single photon source
based on cavity-assisted Raman transition.

\acknowledgments

We acknowledge support from the Alexander von Humboldt Foundation,
and thank G. Giedke and E. Knill for useful discussions.

\end{document}